# Influence of point defects on magnetic vortex structures

M. Rahm[1], R. Höllinger[1], V. Umansky[2], and D. Weiss[1]

[1]Institut für Experimentelle und Angewandte Physik, Universität Regensburg, D-93040 Regensburg, Germany
[2]Braun Center for Submicron Research, Weizmann Institut, Rehovot 76100, Israel

We employed micro-Hall magnetometry and micromagnetic simulations to investigate magnetic vortex pinning at single point defects in individual submicron-sized permalloy disks. Small ferromagnetic particles containing artificial point defects can be fabricated by using an image reversal electron beam lithography process. Corresponding micromagnetic calculations, modeling the defects within the disks as holes, give reasonable agreement between experimental and simulated pinning and depinning field values.

In recent years both fabrication of submicron-sized ferromagnetic particles and measurement techniques to investigate their magnetic properties have been established. Magnetic vortices are prominent representatives of coherent spin structures and occur as ground states in disk-shaped submicron-sized ferromagnetic samples. In order to minimize stray field energy their magnetization lies in the plane of the disk and is aligned along the edge to close the magnetic flux. Towards the center of the vortex, however, the increasing exchange energy density causes the magnetization to turn out of the plane. This leads to the formation of a small region with perpendicular magnetization at the vortex core.[1,2,3] Magnetic vortices have not only been observed in disk-shaped samples but were also reported to play a dominant role in the magnetization reversal of small rectangular and elliptic elements.[4,5,6]

Using micro-Hall magnetometry it was shown recently that the core region of a magnetic vortex can be pinned at a point defect within a ferromagnetic disk.[7] Here, we will present Hall measurements together with simulations to get a more detailed understanding of the pinning process. Especially, the influence of the defect's structure on the interaction between magnetic vortex and defect will be examined. Our investigations show that artificial point defects in ferromagnetic submicron particles control the magnetic behavior. This property might be useful in magnetoelectronic devices.

The micro-Hall sensors were fabricated from GaAs-AlGaAs heterojunctions containing a two-dimensional electron gas (2DEG) at the interface. The low carrier density of about $5 \cdot 10^{11}$ cm$^{-2}$ guarantees a high Hall signal, while the distance between the sample surface and the 2DEG is kept small (here: 35 nm) to increase the effect of the stray field on the Hall signal. The cross-shaped sensor structure with 700-nm-wide bars was defined by electron beam lithography (EBL) and reactive ion etching (using SiCl$_4$). In a next step the ferromagnetic disk needs to be placed on the Hall cross.

To intentionally incorporate defects in a disk we developed a special process using the image reversal properties of PMMA if irradiated by high electron doses during electron beam lithography.[8,9] First, the Hall cross is coated with a double resist system (bottom layer: ~150 nm with 200 ku molecular weight; top layer: ~50 nm with 950 ku) to facilitate lift-off. Then, the designated disk area is defined by EBL using a conventional positive resist process with 25 keV electron energy and doses in the range from 260 to 300 µC/cm$^2$. If a defect is to be incorporated in the disk the corresponding spot is overexposed using an electron dose above 400 fC. While the moderately exposed resist, defining the disk, is removed in the subsequent developer step (using a 3:7 solution of ethylene glycol monoethyl ether in methanol) the overexposed resist spot remains, forming a pillar. After e-beam evaporation of 5 nm Ti as adhesive layer, 30 nm permalloy is deposited by thermal evaporation and covered with 10 nm Ti as a protective layer, before the magnetic disk on the Hall cross is obtained by lift-off in acetone. A completed disk on a Hall cross containing two defects is shown in Fig. 1. The defects, which are not removed by acetone, typically had diameters up to 85 nm and stuck up to 110 nm out of the disk's surface.

Fig. 1 shows that the disk is not placed in the center of the Hall cross. As the magnetic field $H_{ext}$ is applied in the plane of the disk, the Hall signal can be maximized by placing the disk at the border of the active area, which is, in the ballistic transport regime, defined by the junction region of current and voltage bars. The measured Hall voltage is proportional to the magnetic stray field component perpendicular to the 2DEG, averaged over the active area.[10] The experiments were carried out at 1.7 K (ballistic transport regime) in a $^4$He-cryostat containing a superconducting coil to produce $H_{ext}$. The magnetic field, applied in-plane and hence causing no Hall voltage, changes the sample's magnetization pattern and accordingly its stray field. The latter is detected as voltage change in the Hall

Author to whom correspondence should be addressed; electronic mail: michael.rahm@physik.uni-regensburg.de

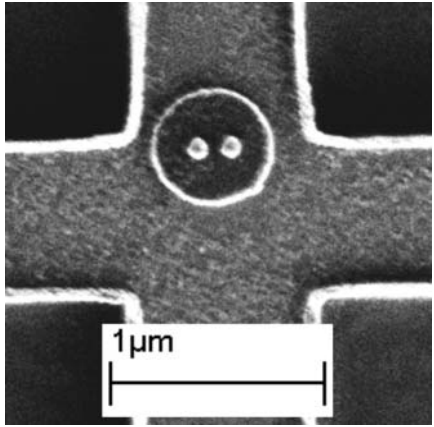

FIG. 1. Permalloy disk (diameter: 500 nm, thickness: 30 nm) with two artificial defects on top of a Hall sensor. The defects have diameters of about 85 nm; their center to center distance is ~150 nm. Detailed investigations of the magnetic behavior of disks containing two or more defects will be published elsewhere.

signal by using conventional lock-in techniques. Previous experiments carried out on disks have shown that the measured Hall voltage maps nearly perfectly the corresponding calculated hysteresis trace.[11]

In-plane magnetization reversal curves obtained by micro-Hall magnetometry from disks without defects are characterized by two distinct jumps at higher $H_{ext}$ and a nearly linear change of the Hall voltage in between.[7] While the jumps are due to nucleation and annihilation of a magnetic vortex, the smooth section stems from the reversible shift of the vortex's center perpendicular to the direction of $H_{ext}$. The gray line in Fig. 2 was measured on a permalloy disk containing a point defect located about 40 nm off the center of the disk (see inset of Fig. 2).[12] The defect affects the hysteresis trace drastically.[7] In the field range between 40 Oe and -360 Oe the curve deviates from one obtained from a disk without defect. In this field range the slope of the curve is reduced and forms a plateau-like structure (interrupted by a tiny jump at ~ -100 Oe, which belongs to a small hysteresis loop described below). The plateau was shown to correspond to a pinned vortex state in Ref. 7. The huge jump of the Hall signal at 40 Oe, which is due to vortex nucleation, ends directly at the plateau. This means that the vortex is already formed in the pinned state. The jump at -360 Oe is caused by release of the vortex core from the point defect. Finally, the step at -750 Oe indicates the transition from the vortex state to saturation. The field range between depinning at -360 Oe and elimination of the vortex at -750 Oe is characterized by a smooth decrease of the measured Hall signal, which is typical for unimpeded movement of the vortex core.

To compare with experiment we performed micromagnetic simulations with LLG Micromagnetics Simulator.[13] In particular, we intended to clarify the role of the magnetic material covering the defect. To this end the disk was discretized in 5 nm cubic cells; typical values for the exchange stiffness (1.3 μerg/cm) and the saturation magnetization (800 emu/cm³) were used. Three different models were explored; the results are plotted in Fig. 2. To compare experiment and simulation we normalized the data to the saturation values of Hall voltage and magnetization, respectively. In the first model we described the defect simply as a hole with a diameter of 25 nm, neglecting its inevitable permalloy covering (model A, solid line in the graph). For the second model, a small magnetic cap was taken into account, which was coupled by dipolar interaction only to the disk (model B, dotted line). The third model allows dipolar and exchange interaction between the cells of the disk and of the magnetic cap (model C, dashed line). Comparing the plateaus ascribed to the pinned vortex state, the first model (hole) fits the measured data best. While for model B the depinning field value is only slightly reduced, the field range of the pinned vortex state is distinctly smaller (0 to -140 Oe) for model C. In contrast to the experiment, where the vortex was formed in the pinned state, in the simulations the vortex was formed in the unpinned state at 60 Oe and was captured at moderately lower fields (at 20 Oe for models A and B and at 0 for model C). Model A provides a good description of our artificial defects, because they are significantly higher than the magnetic material deposited. Indeed, the model was also successfully applied to describe multiple switching of disks containing two or more defects (to be published).

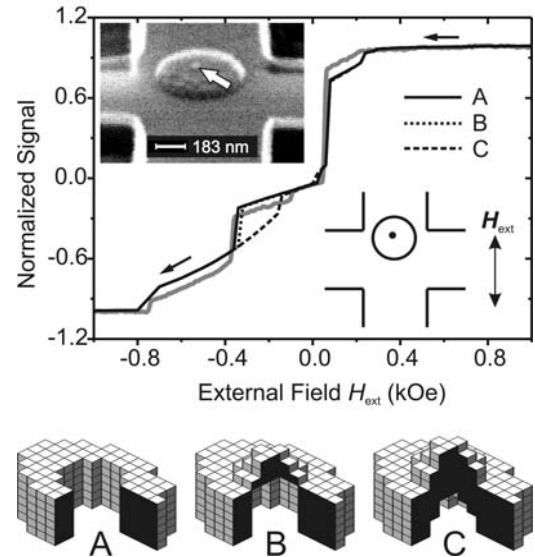

FIG. 2. Comparison of Hall signal and magnetization during reversal. The gray line is the normalized Hall voltage taken from the disk shown in the upper left electron micrograph under a tilt angle of 60°. The disk has a diameter of 500 nm and a thickness of 30 nm; the defect with an approximate diameter of 25 nm is marked by an arrow. The lower right inset sketches the experimental situation, showing the contour of the Hall cross sensor, the boundary of the disk and the defect. The black lines in the graph are calculated, normalized magnetization data, which were simulated by using the different defect models A, B, and C. Three-dimensional images illustrate the discretization grid of the different models A – C in the immediate vicinity of the defect. The size of the cubic cells is 5 nm.

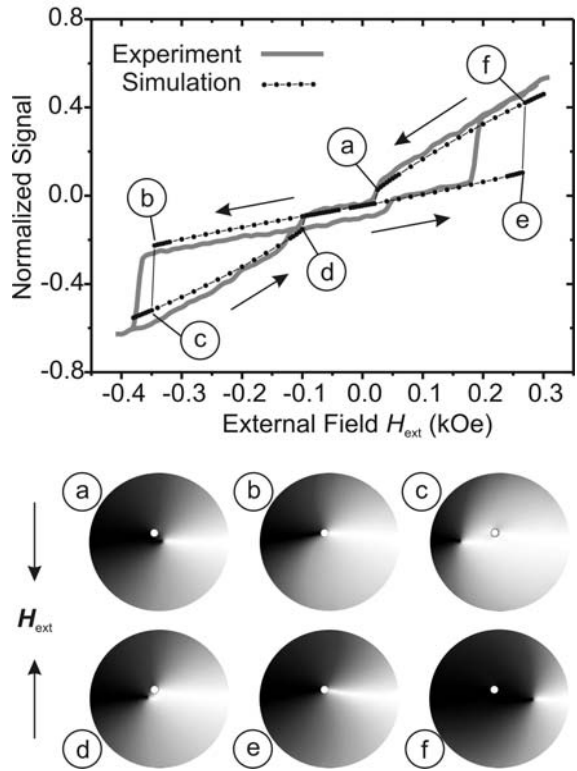

FIG. 3. Minor loops recorded on the same particle as in Fig. 2. The gray and the black lines represent experimental and simulation (model A) data, respectively. The magnetic configuration with a clockwise sense of rotation of the magnetic vortex is shown in the lower part of the figure immediately before pinning at (a) and (d), before depinning at (b) and (e) and immediately after depinning at (c) and (f) for down- and upsweep, respectively. The small measured hysteresis loop with jumps at -100 Oe and 40 Oe could not be reproduced by our simulations.

To compare experimental and simulated pinning and depinning fields we chose model A to calculate minor loops in the field range between -400 Oe and 300 Oe. The resulting traces are depicted in Fig. 3, showing good agreement between experiment and simulation. Fig. 3 covers a smaller range of externally applied fields than Fig. 2, but some of the data shown in Fig. 3 (the down-sweep in the field range of 20 Oe to -400 Oe) perfectly agree with the corresponding data in Fig. 2, as they describe identical situations under unchanged experimental and simulation conditions. The calculated magnetization configurations in the lower part of Fig. 3 illustrate that the small, but nonvanishing magnetic susceptibility in the pinned vortex state is caused by distortion of the vortex state. From (b) to (c) and from (e) to (f) the vortex core is depinned and jumps by about 130 nm off the defect. This process causes the large jumps in experiment and simulation. The asymmetry in the hysteresis trace with respect to the origin is caused by the asymmetric position of the defect on the disk. Although overall shape and calculated pinning and depinning fields agree well with the measured ones, our simulations were unable to grasp the origin of the small hysteresis between -100 Oe and 40 Oe, which appeared as a typical feature in our measurements. Also model B and C were unable to deliver an appropriate explanation. Further investigations will be necessary to clarify the nature of this phenomenon.

In summary, our experiments show that point defects inside ferromagnetic disks can significantly alter the corresponding hysteresis trace. The effects described here might be used for novel switching schemes employing pinning and depinning of vortices. The corresponding micromagnetic calculations are, apart from the small hysteresis close to remanence, in good agreement with experiment.

Support from the DFG (Forschergruppe 370 'Ferromagnet-Halbleiter-Nanostrukturen: Transport, magnetische und elektronische Eigenschaften') is gratefully acknowledged.